\documentclass[draftclsnofoot,onecolumn]{IEEEtran}
\ifCLASSINFOpdf
  \else
  \fi
\usepackage{amssymb}
\usepackage{amsmath}
\usepackage{algorithm}
\usepackage{algorithmic}
\usepackage{graphics}
\usepackage{epsfig}
\usepackage{mathrsfs,}
\usepackage{amssymb}
\usepackage[OT1]{fontenc}
\usepackage[usenames,dvipsnames,svgnames,table]{xcolor}
\usepackage{multirow}

\newtheorem{lm}{Lemma}
\newtheorem{thm}{Theorem}
\newtheorem{cor}{Corollary}

  \usepackage{cite}

\begin{document}

\title{On Solving Ambiguity Resolution with Robust Chinese Remainder Theorem for Multiple Numbers }%with Applications in Sensor Networks}

\author{Hanshen Xiao and Guoqiang Xiao

\thanks{Hanshen Xiao is with CSAIL and the EECS Department, MIT, Cambridge, USA. E-mail: hsxiao@mit.edu.}% <-this % stops a space
\thanks{Guoqiang Xiao is with the College of Computer and Information Science, Southwest University, Chongqing, China. E-mail: gqxiao@swu.edu.cn.}}

%\markboth{IEEE Transactions on Vehicular Technology,~Vol.~XX, No.~XX, XXX~2015}
{}
%{Shell \MakeLowercase{\textit{et al.}}: Bare Demo of IEEEtran.cls for Journals}

\maketitle


\begin{abstract}
Chinese Remainder Theorem (CRT) is a powerful approach to solve ambiguity resolution related problems such as undersampling frequency estimation and phase unwrapping which are widely applied in localization. Recently, the deterministic robust CRT for multiple numbers (RCRTMN) was proposed, which can reconstruct multiple integers with unknown relationship of residue correspondence via generalized CRT and achieves robustness to bounded errors simultaneously. Naturally, RCRTMN sheds light on CRT-based estimation for multiple objectives. In this paper, two open problems arising that how to introduce statistical methods into RCRTMN and deal with arbitrary errors introduced in residues are solved. We propose the extended version of RCRTMN assisted with Maximum Likelihood Estimation (MLE), which can tolerate unrestricted errors and bring considerable improvement in robustness.
\end{abstract}


\IEEEpeerreviewmaketitle

\section{Introduction}
\noindent Localization of nodes \cite{phase1}, \cite{phase2} and frequency estimation \cite{sensor} are two fundamental problems in sensor networks. Due to the restriction on precise synchronization and hardware resources such as high-rate analog to digital converters (ADC), the phase detection based ranging methods and sub-Nyquist sampling are two important approaches used in these kinds of applications. Especially, the radio interferometric positioning system (RIPS) \cite{rips}, which receives considerable attraction recently, is also based on the idea to measure the phase of the interference signals generated by two transmitters. However, all of above-mentioned methods are confronted with the ambiguity resolution problems.

%Since hardware and energy resources are limited in sensor networks, it is hard to achieve precise synchronization and high capacity of computing and storage, which are required to achieve high accuracy localization \cite{phase1}, \cite{phase2} and frequency estimation \cite{sensor} among the nodes. In recent literature, under-sampling approaches are widely used in these kinds of applications. The modeling of these problems has been well studied and can be broadly categorized as the range ambiguity resolution and frequency ambiguity resolution.

To be formal, let $m_l$, $l=1,2,...,L$, denote a group of moduli selected and $X_i$, $i=1,2...,N$, denote multiple numbers. In the model of undersampling frequency estimation \cite{sensor}, \cite{fre2},\cite{fre2017}, \cite{SP2017} $\{X_i\}$ represent the frequencies to be estimated and the moduli $\{m_l\}$ stand for the sampling frequency used. For a complex waveform $f(t) = \sum_{i=1}^{N} A_i e^{2\pi j X_i t}$ sampled in frequency $m_l$, the undersampled waveform becomes $x_{m_l}[n] = \sum_{i=1}^N A_i e^{\frac{2 \pi j X_i n}{m_l}}, n \in \mathbb{Z}$.  The spectrum of $x_{m_l}[n]$ can be obtained via an $m_l$-point Discrete Fourier Transform, i.e., $DFT_{m_l}(x_{m_l}[n]) [k]=\sum_{i=1}^{N}A_i \mathbf{1}(k-\langle X_i \rangle _{m_l})$, where the residue sets, $ \{ r_{il}=\langle X_i \rangle_{m_l} | i=1,2,...,N \}, l=1,2,...,L$, can be read from the peaks on spectrum respectively, though the correspondence between $X_i$ and $r_{il}$ is unknown. Here $\langle X_i \rangle_{m_l}$ denotes the residue of $X_i$ modulo $m_l$ and $\mathbf{1} $ is the indicator function. Similarly, in a localization system \cite{phase1}, \cite{phase2}, $\{X_i\}$ stand for the distances and $\{m_l\}$ denote the wavelengths, respectively. In a nutshell, addressing the ambiguity problems is equivalent to recover $X_i$ with the residue sets, $\{ r_{il}\}$, which is a generalized Chinese Remainder Theorem (CRT) problem. It is well known that CRT describes a closed-form relationship between an integer and its residues modulo given pairwise co-prime moduli. However,  even if very small errors are introduced in the residues, it may result in an incredibly large deviation in reconstruction with conventional CRT. In the presence of error $\Delta_{il}$ in each residue, which is almost inevitable in practice, the problem turns to estimate $X_i$ with $\widetilde{X}_i$ , which is reconstructed by erroneous residues sets, $R_{l} = \{ \widetilde{r}_{il} =\langle X_i + \Delta_{il} \rangle_{m_l}| i=1,2,...,N \}, l=1,2,...,L$.

% In the model of multiple frequency estimation, the moduli represent the undersampling frequency and $X_i$ is the frequency to be estimated. In each sampling with $m_l$, we obtain sets $ \{ r_{il}=\langle X_i \rangle_{m_l} | i=1,2,...,N \}, l=1,2,...,L$, after applying Fourier Transform on undersampled waveforms, where $r_{il} = \langle X_i \rangle_{m_l}$ denotes the residue of $X_i$ modulo $m_l$. Similarly, in the instance of modeling distance measurement, $X_i$ stands for the distances and denote by the wavelength $m_l$, respectively.
%Solving ambiguity resolution problem plays an importment role in Sensor networks  Location System \cite{phase},\cite{phase2} ... Radio interferometric positioning system (RIPS) .. Multiple frequencies determination by multiple sensors  with low sampling rates. \cite{sensor} considers the case of real waveforms, which can be essentially generalized to ... frequency estimation from undersampling waveforms and range ambiguity resolution in phase unwrapping and distance measurement. Some other applications such as distributed storage in sensor networks can also be mathematically modeled to the design of RCRT.

%Statistically estimation can be referred to \cite{musiccrt}. Comparing to the conventional CRT, a pure mathematical issue, the study on RCRT is a newly emerging field and promoted by specific engineering applications such as

 %The reason is essentially that an integer is not proportional to the magnitude of its residues.
 To address the problem arising from error sensibility, Robust CRT (RCRT) is formally proposed and studied during last decade. Ideally, RCRT is expected to achieve a reconstruction deviation proportional to the errors in residues. The studies in this area have been elaborated in \cite{survey}. %It is noted that when $m_l$ is bigger than $X_i$, the $\widetilde{r}_{il}$ should be of $X_i$ itself. Therefore, given residue error bound $\delta = \max_{il} \Delta_{il}$, it is the best performance we can expect that reconstruction error $|\widetilde{X}_i - X_i|$ is no bigger than $\delta$.
Certainly, any improvement over the error bound or the dynamic range $\max_{i} X_i$ in RCRT will lead to more robust and efficient estimation schemes in many applications.   %Now it can be observed that a design of RCRT can straightforwardly lead to a robust frequency estimation. Interested readers can refer to the references for more details about other listed applications.

In this paper, trodding the line of research in \cite{sp}, \cite{mle},\cite{phase1}, we initiate the study on RCRT for multiple numbers (RCRTMN) with tolerance of arbitrary errors. Besides presenting the specific algorithms, we show how to sharply reduce the complexity and introduce MLE for further improvement.
% a novel generalized RCRT for multiple numbers (RCRTMN) with tolerance of arbitrary errors is proposed. Different from \cite{TSP2018} which is deterministic, the extended version method is further assisted with maximum likelihood estimation.

%the generalization is based on the idea that combing both classic remainder code and RCRT in \cite{sp2015} while the method is different.

%Framework: 1. finding out the cutting point

%Similar to the error correction codes, RCRT achieves robustness by sacrificing the dynamic range, i.e., adding redundancy in the representation residue space to tolerate error.


\section{Preliminaries}
\label{}

\subsection{Remainder Codes for Hamming-weight Errors }
\noindent Error correction coding is a well-studied field, where most research has concentrated on errors measured with the Hamming weight. Classic remainder code under such scenario was formally proposed during the 1960s. The first polynomial time error correction scheme was constituted by Goldreich et al. \cite{RCRT} based on LLL lattice reduction and further improved by Guruswam et al. in \cite{RCRT2}. We conclude their results as the following lemma.

\begin{lm}
\label{HMCRT}
Given $L$ co-prime integers, $M_1, M_2, ... , M_L$, which are in an ascending order, the residue vector of an integer $X$ within $[0,\prod_{l=1}^{K}M_l)$, $K \le L$, is expressed as $\mathbf{x} = (x_1, x_2, ..., x_L)=(\langle X \rangle_{M_1}, \langle X \rangle_{M_2}, ..., \langle X \rangle_{M_L}).$ If there exist $\lambda$ many coordinates that are erroneous in $\hat{\mathbf{x}}=(\hat{x}_1, \hat{x}_2, ..., \hat{x}_L)$, i.e., there exist $\lambda$ many indexes, $l \in \{1,2,...,L\}$, such that coordinate-wise  $x_l \not = \hat{x}_l$ and $\lambda \leq \lfloor \frac{L-K}{2} \rfloor$, then $X$ can be uniquely recovered in polynomial time from the residue vector with errors. %The complexity to decode is $O(Lb\log^c b)$, where $b=\sum_{l=1}^{L} (1+\lfloor \log_2 M_l \rfloor)$ and $c$ is a constant.
%Furthermore, errors can be detected if no more than $(L-K)$ many coordinates of $\mathbf{x}$ happen to be erroneous.
\end{lm}

%The idea behind the classic remainder code is the residue class. For a remainder code of $K$ information moduli and $(L-K)$ redundant moduli, the dynamic range is $[0,\prod_{l=1}^{K} M_l)$. For the residue vector of an integer $X \in [0,\prod_{l=1}^{K} M_l)$, if there exist no more than $r$ erroneous residues, the integer $X'$ recovered from those residues with errors should be no less than $\prod_{l=1}^{K} M_l$, since $X'-X$ is the product of no less than $K$ moduli, i.e., no less than $\prod_{l=1}^{K} M_l$. In addition, it is not hard to observe the minimal distance of the remainder code is exactly $(L-K+1)$, which is the hamming distance between the residue vectors of $0$ and $\prod_{l=1}^{K}$.

Before we proceed, we have to stress the fact that when the number of erroneous residues, $\lambda$, is no bigger than $\lfloor \frac{L-K}{2} \rfloor$, then $\mathbf{x}$, the error-free residue vector of $X$, can be uniquely recovered. However, another noteworthy feature is that, when $\lambda > \lfloor \frac{L-K}{2} \rfloor$, we can still use similar scheme to implement error correction, though it is not guaranteed that there exists a unique code, of which the hamming distance to $\hat{\mathbf{x}}$ is no bigger than $\lambda$. It is clear that $\mathbf{x}$ is one of candidates when $\lambda \leq L-K$, i.e., $X$ is possible to be recovered but may not be distinguished due to multiple possible solutions when $\lfloor \frac{L-K}{2} \rfloor< \lambda \leq L-K.$ In coding theory, to find all possible codes within a fixed distance away from the erroneous vector $\hat{\mathbf{x}}$ is called list decoding. In \cite{RCRT2}, Guruswam et al. proved that there still exists polynomial time list decoding scheme of remainder code when $\lambda < L- \sqrt{KL}$. For general $\lambda$, the corresponding results can be refereed in \cite{etri}. We will use the above results in the following proof.

\subsection{Framework of conventional RCRTMN with bounded errors}
\noindent In the case of bounded errors, assume $\delta > \max_{il} \Delta_{il}$. Denote $\mathscr{M}=\{m_1, m_2, ... , m_L\}$ as the moduli selected, where $m_l=\Gamma M_l$. Throughout the paper, we always assume that $M_l$ are co-prime and $\Gamma=4N\delta$. For the erroneous residues, denote the residue sets as $R_l = \{ \widetilde{r}_{il} = \langle X + \Delta_{il} \rangle_{m_l} | i=1,2,...,N\}$, $l=1,2,...,L$. Let $ \widetilde{r}^c_{il}= \langle  \widetilde{r}_{il} \rangle_{\Gamma}$, which are termed as $common~ reminders$. We arrange the set of common remainders $\{ \widetilde{r}^c_{il} \}$ in an ascending order represented as $R = \{\gamma_1, \gamma_2, ... , \gamma_{\kappa}\}$, where $i=1,2, ... ,N$, $l=1,2,...,L$, and $\kappa \leq NL$. It is not hard to observe that the main difficulty to construct RCRTMN arises from the absence of the correspondence between $\{ \widetilde{r}_{il} \}$ and $X_i$, and interference from introduced errors. We first review the techniques in RCRT from the point of macroscopic view. To achieve robustness, the folding number $\lfloor \frac{X_i}{\Gamma} \rfloor$ plays a key role. It is noted that
\begin{equation}
\label{quo}
  \langle \lfloor \frac{X_i}{\Gamma} \rfloor \rangle_{M_l} = \langle \frac{\langle X_i \rangle_{m_l} - \langle X_i \rangle_{\Gamma}}{\Gamma} \rangle_{M_l} = \langle \frac{r_{il}-r^c_{i}}{\Gamma} \rangle_{M_l}
\end{equation}
Unfortunately, we can not trivially replace $r_{il}$ and $r^c_{i}$ by $\widetilde{r}_{il}$ and $\widetilde{r}^c_{il}$ in (\ref{quo}) due to presence of errors. For example, if there exist some $l_1$ and $l_2$ such that $\Gamma \leq {r}^c_{i} + \Delta_{il_1} < 2\Gamma$ and $0\leq {r}^c_{i} + \Delta_{il_2} < \Gamma$, we have $\langle \frac{ \widetilde{r}_{il_1}-\widetilde{r}^c_{il_1}}{\Gamma} \rangle_{M_{l_1}} = \langle \lfloor \frac{X_i}{\Gamma} \rfloor +1 \rangle_{M_{l_1}}$ while $\langle \frac{ \widetilde{r}_{il_2}-\widetilde{r}^c_{il_2}}{\Gamma} \rangle_{M_{l_2}} = \langle \lfloor \frac{X_i}{\Gamma} \rfloor \rangle_{M_{l_2}}.$ However, things are different if the order of index $l$ is known such that $\{r^c_i + \Delta_{il}\}$ are sorted incrementally, as illustrated in Fig.1, where the order is $l_1, l_2, l_3$. Under this situation, combined with the information of $\widetilde{r}^c_{il} = \langle r^c_{i} + \Delta_{il} \rangle_{\Gamma}$ on a circle modulo $\Gamma$ illustrated in Fig.2, we can recover the relative position of $r^c_i + \Delta_{il}$ represented in the axis in Fig.1, if $\max_{il} \Delta_{il} - \min_{il} \Delta_{il} < \Gamma$. In this example, relative positions of $r^c_{i} + \Delta_{il}$ on axis is determined since it can be inferred that %, from Fig.1, the order of the index is $l_1, l_2, l_3$, which is clearly not consist with the order in Fig. 2, $l_2, l_3, l_1$. However, with the knowledge of $\widetilde{r}^c_{il} = \langle r^c_{i} + \Delta_{il} \rangle_{\Gamma}$ in Fig.2 and the order $l_1, l_2, l_3$,
either $-\Gamma < r^c_1 + \Delta_{il_1} <0$ and $0 \leq r^c_1 + \Delta_{il_j}<\Gamma$, or $0 \leq r^c_1 + \Delta_{il_1} <\Gamma$ and $\Gamma \leq r^c_1 + \Delta_{il_j}< 2 \Gamma$, $j=2,3$.  Anyway, the problem raised by residue inconsistence of  $\lfloor \frac{X_i}{\Gamma} \rfloor$ in (\ref{quo}) is solved naturally. %where in the example, $\langle \frac{\widetilde{r}_{il_1}-\widetilde{r}^c_{il_1}}{\Gamma} +1 \rangle_{M_{l_1}}$ and $\langle \frac{\widetilde{r}_{il_j}-\widetilde{r}^c_{il_j}}{\Gamma} \rangle_{M_{l_j}}$, $j=2,3,$ are all residues of  $\lfloor \frac{X_i}{\Gamma} \rfloor$ modulo $m_l$.
The following lemma is refined from \cite{TSP2018}. %and will present the RCRTMN in  \cite{TSP2018} as Algorithm 1.
\begin{lm}
\label{TSP}
If the error $\Delta_{il}$ introduced in each residue satisfies $\max_{il} \Delta_{il} < \delta = \frac{\Gamma}{4N}$, there exists $j_0 \in \{1,2,...,\kappa-1\}$ such that  $\gamma_{j_0+1 } -  \gamma_{j_0} > 2\delta$ or $j_0 = \kappa$ such that  $\gamma_{1} -  \gamma_{\kappa}+\Gamma > 2\delta$. In addition, for each $i \in \{1,2,...,N\}$, the order of $l$ of $\{\hat{r}^c_{il}\}$ defined in (\ref{hat-1}) and (\ref{hat-2}) in Algorithm 1 below is exactly the same as that of $\{r^c_{i} + \Delta_{il}\}$ when both sets $\{\hat{r}^c_{il}\}$ and $\{r^c_{i} + \Delta_{il}\}$ are arranged in an ascending order. \footnote{Said another way, there exist an empty interval ranging from $\gamma_{j_0}$ to $\gamma_{\langle j_0+1 \rangle_{\kappa}}$ on the circle modulo $\Gamma$ with distance at least $2\delta$ clockwise. We can cut the circle at $\gamma_{j_0}$ and stretch it to be a line, where the relative location of $\hat{r}^c_{il}$ on the line is the same as that of $r^c_{i} + \Delta_{il}$ on the axis in Fig.1. }
\end{lm}

We define $\gamma_{{(l)}_{i}} = \widetilde{r}^c_{il'} = \langle r^{c}_{i} + \Delta_{il'} \rangle_{\Gamma}$, $i \in \{1,2,..., N\}$, where ${(l)}_{i}$ is the index such that $\Delta_{il'}$ is the $l^{th}$ smallest error introduced in the residues of $X_{i}$, illustrated in Fig. 3.
\begin{cor}
\label{cutting}
In Lemma \ref{TSP}, $j_0$ should be ${(L)}_{i_0}$ and $\langle j_0+1 \rangle_{\kappa}$ should be ${(1)}_{i_1}$ for some $i_1,  i_0 \in \{1,2,...,N\}$.
\end{cor}

\textbf{Proof.} Revisit the definition of $\{\hat{r}^c_{il}\}$ in (\ref{hat-1}) and (\ref{hat-2}), which is merely a shift on $\{\widetilde{r}^c_{il}\}$. The ascending order of $\{\hat{r}^c_{il}\}$ corresponds to that of $\Delta_{il}$ for each $i$ based on Lemma \ref{TSP}. Thus $\Delta_{i_0l_0}$ corresponding to $\gamma_{j_0}$ or $\widetilde{r}^c_{i_0l_0}$ should be the maximum for some $i_0 \in \{1,2,...,N\}$, since it corresponds to the largest element of $\{\hat{r}^c_{il}\}$. As the first element of $\{\hat{r}^c_{il}\}$ sorted incrementally, $\gamma_{\langle j_0 +1\rangle_{\kappa}}$ or $\widetilde{r}^c_{i_1l_1}$ corresponding to $\Delta_{i_1l_1}$, should be the smallest for some $i_1 \in \{1,2,...,N\}$. Substituting the notations above, ${j_0} = {(L)}_{i_0}$ and  $\langle j_0+1 \rangle = {(1)}_{i_1} $. Done.

\begin{figure}
\centering
\includegraphics[width=2.6 in,height=0.6 in]{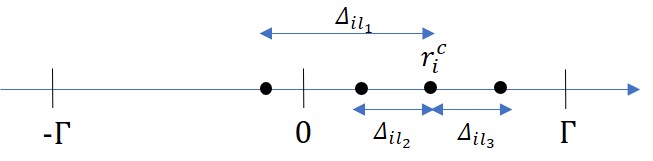}
\caption{Positions of $r^c_i + \Delta_{il}$ on the axis}
\label{Fig1}
\end{figure}

\begin{figure}
\centering
\includegraphics[width=1.71in,height=1.7in]{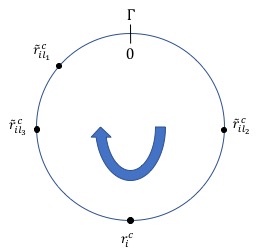}
\caption{Positions of $\widetilde{r}^c_{il} = \langle r^c_i + \Delta_{il} \rangle_{\Gamma}$ on the circle modulo $\Gamma$.}
\label{circle}
\end{figure}

The rest work is to apply Generalized CRT for multiple numbers (GCRTMN) reconstruction in the error-free case \cite{TSP2018} on $\widetilde{q}_{il} = \langle  \frac{ \widetilde{r}_{il}- \hat{r}^c_{il}}{\Gamma}  \rangle_{M_l}$ to uniquely recover $\widetilde{q}_{i}$ for each $i$, where $\langle \widetilde{q}_{i} \rangle_{M_l} =\widetilde{q}_{il}$. The reconstruction of  $\widetilde{q}_{i}$ also naturally determines the correspondence between $X_i$ and $\widetilde{r}^c_{il}$. To deal with bounded errors, no redundant moduli are required. Finally, we conclude the algorithms in \cite{TSP2018} as follows with $K$ moduli.

 \begin{algorithm}
\caption{RCRTMN with bounded errors in \cite{TSP2018}}
\label{alg:RCRT-arbi}

\textbf{Input.} Moduli: $\{m_1=\Gamma M_1, m_2=\Gamma M_2, ... , m_K=\Gamma M_K\}$. Residue Sets: $R_l = \{ \widetilde{r}_{il} | i=1,2,...,N\}$, $l=1,2,...,K$.
\begin{itemize}

\item Step 1. Calculate the common residues $\gamma_j = \langle \widetilde{r}_{il} \rangle_{\Gamma} $, $j=1,2,...,\kappa$, arranged in an ascending order .

\item Step 2. Find out $j_0 \in \{1,2,...,\kappa-1\}$ such that  $\gamma_{j_0+1 } -  \gamma_{j_0} > 2\delta$ or $j_0 = \kappa$ such that  $\gamma_{1} -  \gamma_{\kappa}+\Gamma > 2\delta$

\item Step 3. When $j_0 \not = \kappa$, for each $i$ and $l$, if $ \widetilde{r}^c_{il}  > \gamma_{j_0}$, define
\begin{equation}
\label{hat-1}
%\hat{r}_{il} =\widetilde{r}_{il}+\Gamma; ~~
\hat{r}^c_{il}= \widetilde{r}_{il} - \Gamma.
\end{equation}
else if $  \widetilde{r}^c_{il} \leq \gamma_{j_0}$,
\begin{equation}
\label{hat-2}
%\hat{r}_{il} =\widetilde{r}_{il}; ~~
 \hat{r}^c_{il}= \widetilde{r}_{il}.
\end{equation}
When $j_0 = \kappa$, $\hat{r}^c_{il}= \widetilde{r}_{il}$ for each $i$ and $l$.
\item Step 4.  Let
\begin{equation}
\label{q_ij}
\widetilde{q}_{il} = \langle  \frac{ \widetilde{r}_{il}- \hat{r}^c_{il}}{\Gamma}  \rangle_{M_l}
\end{equation}
and apply GCRTMN on $\widetilde{q}_{il}$ to recover $\widetilde{q}_i$, $i=1,2,...,N,$ $\widetilde{q}_{il} = \langle \widetilde{q}_i \rangle_{M_l}$, and correspondence between $\widetilde{X}_i$ and $\widetilde{r}_{il}$.
\end{itemize}

\textbf{Output.} $\widetilde{X}_{i} = \Gamma \widetilde{q}_i + [\frac{ \sum_{l=1}^{K} \hat{r}^c_{il}}{K}]$, $|\widetilde{X}_{i}-X_i| < \delta$, as the estimation of $X_i$, where $[*]$ is the round operation of $* \in \mathbb{R}$.

\end{algorithm}

%due to the error detection ability mentioned, a naive way to remove the unbounded errors is to enumerate all combination of $\lfloor \frac{L-K}{2} \rfloor$ many $R_l$ and figure out whether the group of $R_l$ collected is without unbounded errors. Clearly it will result in an exponential complexity, i.e., $O(\binom{L}{k})$. In the following, we propose a polynomial time algorithm to solve the problem effectively.

\section{RCRTMN with Arbitrary Errors}
\noindent We divide the construction of RCRTMN for arbitrary errors into two parts, the estimation of folding number $\lfloor \frac{X_i}{\Gamma} \rfloor$ and the common residues $\langle X_i \rangle_{\Gamma}$, respectively. Throughout the rest of the paper we consider a system formed by $L$ moduli, where $(L-K)$ moduli are redundant. We merely assume that $\prod_{l=1}^{K} M_l$ is big enough, where the specific lower bound of $\prod_{l=1}^{K} M_l$ given $X_i$ to apply GCRTMN on $\widetilde{q}_{il}$ to uniquely recover $\widetilde{q}_{i}$  can be referred in \cite{TSP2018}. Based on Lemma \ref{HMCRT}, we can correct up to $\lfloor \frac{L-K}{2} \rfloor$ errors. Following the notations given before, for each $\gamma_j$, $j \in \{1,2,...,\kappa\}$, let $I_j$ denote the interval $(\gamma_j,\gamma_j+2\delta]$, if $\gamma_j+2\delta<\Gamma$; otherwise, $I_j=(\gamma_j,\Gamma) \cup [0,\gamma_j+2\delta-\Gamma]$. For each $\gamma_j=\widetilde{r}^c_{il}$, it is assigned with a label $[l]$, the index of the residue set it belongs to. In the following, we divide the index set $ \{1,2,...,L\}$ into two parts $\mathscr{G}$ and $\mathscr{B}$: $R_l$, $l \in \mathscr{G}$, are called $Good$ residue sets, in which the errors are bounded by $\delta$; while $R_l$, $l \in \mathscr{B}$, are called $Bad$ residue sets, in which the errors can be arbitrary and unbounded. Throughout the paper, we always assume $| \mathscr{B} | \leq \lfloor \frac{L-K}{2} \rfloor$. Let $\mathscr{R}_{j}$ denote the set of labels in $I_{j}$. For example, in Fig.4, $\mathscr{R}_{{(1)}_{i_1}}$ for the interval $I_{{(1)}_{i_1}}$ is $\{ [l'_2], [l'_3] \}$. Let $\mathscr{N}$ denote the index set for those $j$ such that, for $I_j$, $ |\mathscr{R}_j| \leq \lfloor \frac{L-K}{2} \rfloor$ is satisfied and the label of $\gamma_j$ is not in $\mathscr{R}_j$.

\subsection{Folding Number Estimation}
\noindent In the presence of arbitrary errors, Lemma 2 and Corollary 1 both are no longer tenable since there may exist an $\widetilde{r}^c_{il}$, $l \in \mathscr{B}$, between $\widetilde{r}^c_{i_0l_0}$ and $\widetilde{r}^c_{i_1l_1}$ on the circle modulo $\Gamma$. Neverless, if there does exits an empty $I_{j}$, then $j$ here can certainly be such $j_0$ in Lemma 2 and with the same definition on $\hat{r}^c_{il}$ and $\widetilde{q}_{il}$, the problem is easy to solve. In order to find two successive elements in $\{\widetilde{r}^c_{il}, l \in \mathscr{G} \}$ with distance at least $2\delta$ over the circle, we will construct such an interval without any $\widetilde{r}^c_{il}$. To this end, for an $I_{t}$, $t \in \mathscr{N}$, we remove all residues, $\gamma_{j}$, where $\gamma_j = \widetilde{r}^c_{il}, l \in \mathscr{R}_t$. Let $\{\gamma'_{j}\}$ denote the left $\kappa'$ many common residues in an ascending order. Since the label of $\gamma_t$ is not within $\mathscr{R}_t$, $\gamma_t$ will be kept and denote $\gamma_{t}$ as $\gamma'_{t'}$ in $\{\gamma'_{j}\}$. Let $\mathscr{G'} = \mathscr{G} \cap \bar{\mathscr{R}_t} $ and $\mathscr{B'} = \mathscr{B} \cap \bar{\mathscr{R}_t}$, where $\bar{\mathscr{R}_t}= \{1,2,...,L\} / \mathscr{R}_t$. Since the number of residue sets removed, $|\mathscr{R}_t|$, is upper bounded by $\lfloor \frac{L-K}{2} \rfloor$, thus $|\mathscr{G'}| \geq |\mathscr{G}| - |\mathscr{R}_t| \geq K$. We will show in the following that we can always find such an $I_{t}$ on the circle and consequently, by applying Lemma 2 on the rest residues, $\widetilde{q}_i$ can be uniquely recovered similar to Algorithm 1 with list decoding of errors in Hamming weights up to $\lfloor \frac{L-K}{2} \rfloor$.

%there is neither guarantee nor distinguishability to find out two successive $\widetilde{r}^c_{i_0l_0}$ and $\widetilde{r}^c_{i_1l_1}$ with distance clockwise at least $2\delta$ on the circle modulo $\Gamma$, where $l_0, l_1 \in \mathscr{G}$.

\begin{lm}
\label{main}
For an $I_{t}$, $t \in \mathscr{N}$, let $\gamma'_{t'} = \gamma_t$ after residues are removed. In the case of $t' \not = \kappa'$, let $\hat{r}^c_{il} = \widetilde{r}^c_{il} - \Gamma$, when $\widetilde{r}^c_{il} = \langle \widetilde{r}_{il} \rangle_{\Gamma} > \gamma_{t_0}$; let $\hat{r}^c_{il} =\widetilde{r}^c_{il}$, when $\hat{r}^c_{il} \leq \gamma_{t_0}$. In the case of $t' = \kappa'$, let $\hat{r}^c_{il} =\widetilde{r}^c_{il}$. Then for each $i \in \{1,2,...,N\}$ and $l \in \mathscr{G'}$, the relative location of $\hat{r}^c_{il}$ is exactly the same as that of ${r}^c_{i}+\Delta_{il}$ on axis.
  \end{lm}

\textbf{Proof.} With the notations above, now we can find two successive $\gamma'_{t'}$ and $\gamma'_{\langle t'+1 \rangle_{\kappa'}}$ such that $$\gamma'_{\langle t'+1 \rangle_{\kappa'}} - \gamma'_{t'} + \textbf{1}(t' = \kappa') > 2\delta,$$ where $\textbf{1}(t' = \kappa') =1$ iff $t' = \kappa'$, otherwise it equals 0. We search two elements $\widetilde{r}^c_{il}$, $l \in \mathscr{G'}$, say $\gamma'_{\alpha}$ and $\gamma'_{\beta}$, closest to $\gamma'_{t'}$ counterclockwise and to $\gamma'_{\langle t' +1\rangle_{\kappa'}}$ clockwise, respectively. Then $\hat{r}^c_{il}$, $l \in \mathscr{G'}$, defined in Lemma 3 is the same as those obtained in (\ref{hat-1}) and (\ref{hat-2}) when $j_0 = \alpha$. Furthermore, the clockwise distance between $\gamma_{\alpha}$ and $\gamma_{\beta}$ is at least $2\delta$. Thus based on Lemma \ref{TSP}, for $l \in \mathscr{G'}$, the claim holds and the relative location of $r^c_i + \Delta_{il}$ on axis is determined. Done.
\begin{figure}
\centering
\includegraphics[width=3.3 in,height=0.55 in]{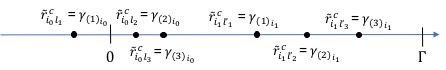}
\caption{An example for positions of $r^c_i + \Delta_{il}$ on the axis where $N=2$ and $L=3$.}
\label{Fig1}
\end{figure}
\begin{figure}
\centering
\includegraphics[width=2.31 in,height=1.7 in]{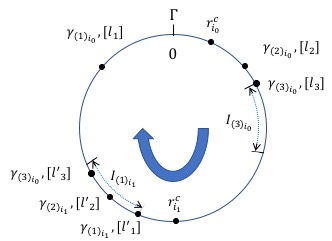}
\caption{Continuing example for positions of  $\widetilde{r}^c_{il}$($\gamma_j$) on the circle modulo $\Gamma$.}
\label{circle}
\end{figure}
% If the number of labels contained in $\lfloor \frac{L-K}{2} \rfloor$ is no bigger than $\lfloor \frac{L-K}{2} \rfloor$, then we delete all the $\gamma_j$ with the same index as those in $I_j$. For the rest, we select $\gamma_{j'}$ as the smallest element in

With Lemma \ref{main}, %the worst case for $\hat{r}_{il} =\widetilde{r}_{il}$ is that $\mathscr{G}=L-2 \lfloor \frac{L-K}{2} \rfloor$ and $| \mathscr{B} | = \lfloor \frac{L-K}{2} \rfloor$.
we can similarly define $\widetilde{q}_{il}= \langle \frac{ \widetilde{r}_{il} - \hat{r}^c_{il}}{\Gamma} \rangle_{M_l}$. After removing all residue sets with same labels in $I_{t}$, $t \in \mathscr{N}$, there exist at most $\lfloor \frac{L-K}{2} \rfloor$ many erroneous $\widetilde{q}_{il}$, $l \in \mathscr{B'}$, which are not the residue of $\widetilde{q}_i$. Here $\widetilde{q}_i$ is the same notation as Algorithm 1 and satisfies $\widetilde{q}_{il} = \langle \widetilde{q}_i \rangle_{M_l}$ for $l \in \mathscr{G'}$. However, in the worst case that all labels of $\gamma_{j}$ in $I_{t}$ are from $\mathscr{G}$ and $| \mathscr{B} | = |\mathscr{R}_t| = \lfloor \frac{L-K}{2} \rfloor$, after residue sets removed, it is reduced to a system with $L- \lfloor \frac{L-K}{2} \rfloor$ moduli, where $| \mathscr{B} | = | \mathscr{B'} | = \lfloor \frac{L-K}{2} \rfloor$ and $| \mathscr{G'} | = L-2\lfloor \frac{L-K}{2} \rfloor$. As there are merely $L-K-\lfloor \frac{L-K}{2} \rfloor$ redundant moduli left, based on Lemma 1, the number of errors exceeds the unique correction capability $\lfloor( L-K- \lfloor \frac{L-K}{2}\rfloor )/2 \rfloor$.  When we apply list decoding \cite{etri}, \cite{RCRT2} to correct up to $\lfloor \frac{L-K}{2} \rfloor$ errors in each step of GCRTMN \cite{TSP2018} on $\widetilde{q}_{il}$, it is not guaranteed that $\widetilde{q}_i$ can be uniquely recovered from $\widetilde{q}_{il}$. Nevertheless, $\widetilde{q}_i$ should be in the decoding list since $|\mathscr{G'}| \geq K$ and $| \mathscr{B'} | \leq  \lfloor \frac{L-K}{2} \rfloor$. On the other hand, based on Lemma 2, there exists $\gamma_{j_0} = \widetilde{r}^c_{il}$, $l \in \mathscr{G}$, such that $I_{j_0}$ does not contain any $\widetilde{r}^c_{il}$, $l \in \mathscr{G}$. Therefore, $j_0 \in \mathscr{N}$ and the labels of elements in $I_{j_0}$ must be all from $\mathscr{B}$ if they exist.
%when set $\{\widetilde{r}^c_{il}\}$, $l \in \mathscr{G}$ and $i=1,2,...,N$, is arranged in an ascending order, there exist two successive $\widetilde{r}^c_{il}$ with their distance at least $2\delta$, where we denote the pair as ($\gamma_{\zeta}, \gamma_{\langle \zeta+1 \rangle_{\kappa}}$). Even if the unbounded errors introduced, $I_{zeta}$ is a candidate satisfying the conditions in Lemma 3. The labels in $I_{zeta}$ must be all come from $\mathscr{B}$.
  Assuming that $| \mathscr{R}_{j_0}|$ is $\tau$, then the number of residue sets or moduli left is $L-\tau$ and $|\mathscr{B'|} = \lfloor \frac{L-K}{2} \rfloor-\tau$. Therefore the error correction capacity is  $\lfloor \frac{L-\tau-K}{2} \rfloor$, which is no less than $\lfloor \frac{L-K}{2} \rfloor-\tau$. Thus we enumerate the operation on each $I_t$, $t \in \mathscr{N}$, with list decoding based error correction until $\widetilde{q}_{i}$ can be distinguished with the unique solution. We formally conclude the scheme as follows. %Without loss of generality, we assume that for $l=1,2, ... , K+r$, errors in residue sets $R_l$ are all bounded by $\frac{\Gamma}{4N}$.  However, for $l=K+r+1, K+r+2, ... , L$, there still exist unbounded errors in $\hat{R}_l$. Via the proof in \cite{TSP2018}, if there is no unbounded errors, i.e., the error bound is $\frac{\Gamma}{4N}$ for errors in each residue, then there exists two successive common residues in an ascending order such that $\gamma_{\langle j+1 \rangle} - \gamma_{j}+1_{j=\kappa} \Gamma > 2\delta$. Therefore, if what we remove in step 3 of the algorithm have less than select the pair $\gamma_j$ and $\gamma_{j'}$, where $\gamma_{j'}$ is the smallest residue outside the ...

\begin{algorithm}
\caption{RCRTMN for arbitary errors}
\label{alg:RCRT-arbi}
\textbf{Input.} Moduli: $\{m_1=\Gamma M_1, m_2=\Gamma M_2, ... , m_L=\Gamma M_L\}$. Residue Sets: $R_l = \{ \widetilde{r}_{il} | i=1,2,...,N\}$, $l=1,2,...,L$.
\begin{itemize}
\item Step 1. Calculate the common residues $\gamma_j = \langle \widetilde{r}_{il} \rangle_{\Gamma} $, $j=1,2,...,\kappa$, arranged in an ascending order.

\item Step 2. For each $t \in \mathscr{N}$, do the following steps.

\item Step 3. Delete all the residues with the same labels in $I_{t}$.

\item Step 4. For the rest $\kappa'$ many residues $\gamma'_j$, $\gamma_t = \gamma'_{t'}$. Case 1: $t' \not = \kappa'$. When  $\widetilde{r}^c_{il} > \gamma_{t}$, define $\hat{r}^c_{il} =\widetilde{r}^c_{il}-\Gamma$; Otherwise, $\hat{r}^c_{il} =\widetilde{r}^c_{il}$. Case 2: ${t'}=\kappa'$. Let $\hat{r}^c_{il} =\widetilde{r}^c_{il}$.

\item Step 5. Calculate $\widetilde{q}_{il}= \langle \lfloor \frac{\widetilde{r}_{il} - \hat{r}^c_{il}}{\Gamma} \rfloor \rangle_{M_l}$ and apply Generalized CRT with list decoding based error corrections on each step for $\widetilde{q}_{il}$ to obtain $\widetilde{q}_{i}$, $i=1,2,...,N$.

\item Step 6. For each error correction step, if the solution is unique, output $\widetilde{X}_{i}= \widetilde{q}_{i}\Gamma+[\frac{\sum_{l=1}^{L-|\mathscr{R}_t|} \hat{r}^c_{il}}{L-|\mathscr{R}_t|}$].
\end{itemize}
\end{algorithm}

In section II, we give the notation $(j)_i$. In the rest of the paper, let $\gamma_{(j)_i} = \widetilde{r}^c_{il}$ only for $l \in \mathscr{G}$. It is clear that $(j)_i \not \in \mathscr{N}$ for $j=1,2,...,K-1$, since $\gamma_{(K)_i}, \gamma_{(K+1)_i}, ... \gamma_{(K+\lfloor \frac{L-K}{2} \rfloor)_i}$ are all within $I_{(j)_i}$, i.e., there exist at least ($\lfloor \frac{L-K}{2} \rfloor+1$) labels in such $I_{(j)_i}$ for $j=1,2,...,K-1$. Thus $| \mathscr{N} | \leq (L-K+1)N$ and the complexity of Algorithm 2 is upper bounded by $(L-K+1)N$ times using GCRTMN to recover integers.

In the following, we  proceed to present further optimization to reduce the complexity of Algorithm 2 to $N$ times accessing GCRTMN. Let $\mathscr{P}$ denote the index set for those $j$  such that the number of labels in $I_j$ is no less than $\lfloor K+\frac{L-K}{2} \rfloor$. %, $j \in \mathscr{P}$. % and $\mathscr{H}$ as the index set within which  $j$ such that $I_j$ does not satisfy condition (\ref{condi}).
$\zeta$ and $\zeta'$ in $\mathscr{P}$ are called consecutive index if $\gamma_{\zeta} \in I_{\zeta'}$ or $\gamma_{\zeta'} \in I_{\zeta}$.

%of which the number of labels is no bigger than $\lfloor \frac{L-K}{2} \rfloor$, since there exist at least $L- \lfloor \frac{L-K}{2} \rfloor- j$ labels within $I_{(j)_i}$ for $j=1,2,...,K$. Hence the number of $I_j$ satisfies (\ref{condi}), i.e., $|\mathscr{Q}|$ is upper bounded by $(L-K)N$, and the complexity is at most $(L-K)N$ times access to the GCRT oracle with error correction.

%In most cases, $2K>L$, otherwise there are the number of redundant moduli is not less than that of information moduli. In the following, we assume that $2K>L$.

\begin{thm}
\label{main2}
$\mathscr{P}$ can be divided into at most $N$ disjoint subsets, within which the index are consecutive. Moreover, in Step 2 of algorithm \ref{alg:RCRT-arbi}, $j$ only needs to enumerate the element in $\mathscr{N}$, which is clockwise closet to the first element of a subset.
\end{thm}

%1. At least $K$ Successive indexes in $\mathscr{P}$

%2. There exist one that at least  $L-\lfloor \frac{L-K}{2} \rfloor$ labels in the $I_j$

 \textbf{Proof.}  Clearly, for each $i \in \{1,2,...,N\}$, ${(1)}_i \in \mathscr{P}.$ We claim that for each subset in $\mathscr{P}$, at least one ${(1)}_i$ should be within it. If the claim is true, then the number of such subsets is upper bounded by $N$. Assume there exists a $\zeta \in \mathscr{P}$, $\zeta \not = (1)_i $, which is not successive to $(1)_i $ for any $i$. Since $I_{\zeta}$ contains at least $K+\lfloor \frac{L-K}{2} \rfloor$ labels, which is much bigger than $\lfloor \frac{L-K}{2} \rfloor$, it must contain some labels from $\mathscr{G}$. Supposing $\widetilde{r}^c_{i'l'} \in I_{\zeta}$, $l' \in \mathscr{G}$, then $\gamma_{\zeta} \in I_{(1)_{i'}}$ or $\gamma_{(1)_{i'}} \in I_{\zeta}$, which leads to a contradiction. Next we prove the rest half of the theorem. Recalling Corollary 1 and Lemma 3, there exists $j_0 \in \mathscr{N}$  and clearly $j_0 \not \in \mathscr{P}$ such that $\gamma_{j_0} = \gamma_{(|\mathscr{G}|)_{i_0}}$. Moreover, after removing all $\widetilde{r}^c_{il}$, $l \in \mathscr{R}_{j_0}$, there exists $\gamma_{(1)_{i_1}} $, which is closest to $\gamma_{j_0}$ counterclockwise for all $\widetilde{r}^c_{il}$, $l \in \mathscr{G'}$. Therefore, $\gamma_{j_0}$ is counterclockwise before the first element, denoted by $\gamma_{\zeta'}$, of the consecutive subset containing $\gamma_{(1)_{i_1}}$. Anyway, $\gamma_{\zeta'}$ is lying in the interval ranging from $\gamma_{j_0}$ to $\gamma_{(1)_{i_1}}$ clockwise. In Algorithm 2, when we set $t=j_0$, $\widetilde{q}_{il}$ can be uniquely recovered. On the other hand, it is clear that $\hat{r}^c_{il}$ defined in Step 4 of Algorithm 2 keeps the same for all $l \in \mathscr{G'}$ when we set either $t=\zeta$ or $t=j_0$ in Step 2 of Algorithm 2. Done.

%Recalling the notation $\gamma'_{j'}$ corresponding to $\gamma_{j}$, $j \in \mathscr{G}$, based on Lemma \ref{TSP} and Corollary \ref{cutting}, there exists $\gamma'_{\langle j'+1 \rangle_{\kappa'}} \not \in I_{j'}$, $j' \in \mathscr{G}$ and there exists some $i_0 \in \{1,2,...,N\}$ such that $\gamma'_{\langle j'+1 \rangle_{\kappa'}}  = \gamma_{(1)_{i_0}}$. Thus the start of the successive subset which contains $\gamma_{(1)_{i_0}}$ is clock-wisely after $\gamma_{j'}$,. Done.

 % On the other hand, there must exist some $i'_0 \in \{1,2,...,N\}$ such that $\gamma_{(k+\lfloor \frac{L-K}{2} \rfloor})_{i'_0}$ closet clock-wisely before  $\gamma_{(1)_{i_0}}$. Thu

 % Especially, $(1)_i$ must not be a candidate. For each $i$, we can find the closet clock-wisely $(1)_i'$ comparing to $(|\mathscr{G}|-\lfloor \frac{L-K}{2} \rfloor+1)_i$. Therefore, any $j \in \mathscr{P}$ should satisfy $\gamma_{j} \in \cup_{i \in \{1,2, ... ,N\}} F_i$, where $F_i$ denote the interval from $(|\mathscr{G}|-\lfloor \frac{L-K}{2} \rfloor)_i$ to $(1)_i'$ in the modulo circle. Therefore

Based on Theorem \ref{main2}, the complexity of Algorithm 2 is reduced to $N$ times accessing GCRTMN. Especially when $N=1$, the complexity in \cite{sp} is $\lfloor \frac{L-K}{2} \rfloor$ times higher than that of ours. Moreover, the analysis above is based on reconstruction of multiple integers, but it can be generalized trivially to the real number case \cite{mle}.

%Now we turn back to the proposed algorithm. Based on our assumption that no more than $\lfloor \frac{L-K}{2} \rfloor$ residue sets with errors.

% In step 3, after removing no bigger than $\lfloor \frac{L-K}{2} \rfloor$ residues, there still exist $L-\lfloor \frac{L-K}{2} \rfloor \geq \lfloor \frac{L-K}{2} \rfloor $ redundant moduli. In other words, we can still distinguish whether there exsits other residues with errors by checking whether the reconstructed $\widetilde{q}_{i}$ exceeding the given dynamic range.

 %Specially, for the case that the symmetirc-polynomial based GCRT is applicable, it can tell in the first step to compute the sum of $\widetilde{q}_{i}$. If the reconstructed sum is no less than $\prod_{l=1}^{K} m_l$, then there must exists the removed

\subsection{ Maximum Likelihood Estimation Based Common Residue Estimation }
 \noindent  In Algorithm 2, we briefly give an estimation of $\hat{r}^c_i = X_i - \Gamma \widetilde{q}_i$ by the average of $\hat{r}^c_{il}$, while it is not the maximum-likelihood estimation (MLE). The residue errors, $\{\Delta_{il}\}$, are random variables and may have different variances due to different sampling frequencies in practice. In the following, it is assumed that, for a given $i$, $\{\Delta_{il}\}$ are in wrapped normal distribution with mean 0 and a variance $\sigma_{l}$ for $l=1,2,...,L$, separately. In \cite{mle}, a generic framework on MLE based RCRT for one integer is proposed. Following the idea, we proceed to introduce MLE in our scenario for multiple integers.

    Assume that after recovering $\widetilde{q}_i$, the correspondence between $\widetilde{q}_i$ and $\widetilde{q}_{il}$ for each $l \in \{1,2,...,L\} / \mathscr{R}_t= \bar{\mathscr{R}_t}$ is determined, which further yields the correspondence between $X_i$ and $\widetilde{r}_{il}$. Therefore, the left work is to estimate each $\langle X_i \rangle_{\Gamma}$ separately. According to  \cite{mle}, the MLE of $\widetilde{r}^c_{i}$ is
    \begin{equation}
    \label{mle}
   \mathcal{MLE}({{\widetilde{r}^c}_{i}}) = \arg \min_{0 \leq x < \Gamma} \sum_{l \in \bar{\mathscr{R}_t}}  \frac{1}{{\sigma_l}^2} {d_{\Gamma}}^2 (\widetilde{r}^c_{il},x)
    \end{equation}
    where $d_{\Gamma} (X, Y) = \min_{z \in \mathbb{Z}} | X-Y+z\Gamma |$, i.e., the minimal distance between the residues of $\langle X \rangle_{\Gamma}$ and $\langle Y \rangle_{\Gamma}$ over the circle modulo $\Gamma$.
    In \cite{mle}, it proved that there are $|\bar{\mathscr{R}_t}|$ candidates which can be the optimal solution of (\ref{mle}). In the following, we will derive a simpler  closed-form MLE of $\hat{r}^c_{i}$ in our case. With the assumption of $\Delta_{il}$ given at the start of Section III, since the relative position of elements in $\{\hat{r}^c_{il} \}$ is proved to be the same as that in $\{r^c_{i} + \Delta_{il}\}$, for any $l_1, l_2 \in \mathscr{G'}$, $$ |  \hat{r}^c_{il_1} - \hat{r}^c_{il_2} | \leq |\Delta_{il_1}| + | \Delta_{il_2} | \leq 2\delta. $$ In the following, we assume $2K > L$, i.e., the number of redundant moduli is smaller than that of information moduli. It is noted that $| \mathscr{G'} | \geq L-2 \lfloor \frac{L-K}{2} \rfloor \geq K$. There exists $\min_{l \in \mathscr{G'}} \hat{r}^c_{il} = \hat{r}^c_{il_0}  \in [-\Gamma, \Gamma]$,  such that at least $K$ out of $L$ elements in $\{\hat{r}^c_{il}\}$ are within $[\hat{r}^c_{il_0}, \hat{r}^c_{il_0}+2\delta]$. However, if there exists $\hat{r}^c_{il'}$, which does not belong to any interval $[y, y+2\delta]$, $y \in [-\Gamma, \Gamma]$, which contains at least $K$ many $\hat{r}^c_{il}$, then clearly $l' \not \in \mathscr{G}$. After removing such $\hat{r}^c_{il'}$, denote the set of label $l$ of the rest residues as $\mathscr{H}$, we will show that:

\begin{lm}
\label{range}
$\max_{l_1, l_2 \in \mathscr{H}}  | \hat{r}^c_{il_1} - \hat{r}^c_{il_2} | \leq 4\delta$
\end{lm}

\textbf{Proof.} Assuming the smallest element in $\{\hat{r}^c_{il}\}$, $l \in \mathscr{H}$, is just $\hat{r}^c_{il_1}$, then there are at least $K$ elements $\hat{r}^c_{il}$, $l \in \mathscr{H}$, within $[\hat{r}^c_{il_1}, \hat{r}^c_{il_1}+2\delta]$. Otherwise, there is no interval $[y,y+2\delta]$, $y \leq \hat{r}^c_{il_1}$, that contains both $\hat{r}^c_{il_1}$ and at least other $K-1$ many $\hat{r}^c_{il}$. On the other hand, the number of rest $\hat{r}^c_{il}$, $l \in \mathscr{H}$, beyond $[{\hat{r}^c_{il_1}},{\hat{r}^c_{il_1}}+2\delta]$, is at most $L-K$, which is smaller than $K$. Hence, the rest residues should all be within $[\hat{r}^c_{il_1}+2\delta, \hat{r}^c_{il_1}+4\delta]$. Done. %Therefore $[\hat{r}^c_{il_1} , \hat{r}^c_{il_1}+4\delta]$ includes all $\hat{r}^c_{il}$, $l \in \mathscr{H}$. The claim follows. Done.

From the above lemma, it also indicates that the reconstruction error is upper bounded by $3\delta$. Especially, when $N=1$, i.e., $\Gamma = 4\delta$, after removing all $\hat{r}^c_{il}$ with the same label as those in an $I_t$ of length $2\delta$, the rest $\hat{r}^c_{il}$ are all within an interval no bigger than $\Gamma - 2\delta = 2\delta$. Substituting $\Gamma= 4N \delta$, $\max_{l_1, l_2 \in \mathscr{H}}  | \hat{r}^c_{il_1} - \hat{r}^c_{il_2} | \leq \frac{\Gamma}{N} \leq \frac{\Gamma}{2} $, when $ N \geq 2$. Moreover, noticing that $d_{\Gamma}(X,Y) \leq \frac{\Gamma}{2}$, therefore,
    $ |\hat{r}^c_{il_1}-\hat{r}^c_{il_2}| = d_{\Gamma} ( \hat{r}^c_{il_1}, \hat{r}^c_{il_2}) =  d_{\Gamma}( \widetilde{r}^c_{il_1}, \widetilde{r}^c_{il_2}) \leq \frac{\Gamma}{2}$
     for any $l_1, l_2 \in \mathscr{H}$, as $\Gamma | (\widetilde{r}^c_{il}-\hat{r}^c_{il})$. Therefore referring to (\ref{mle}), the MLE of $\hat{r}^c_{il}$, $l \in \mathscr{H}$, can be expressed as
      \begin{equation}
    \label{mle1}
   \mathcal{MLE}({{\hat{r}^c}_{i}}) = \arg \min_{\min \hat{r}^c_{il}  \leq x \leq \max \hat{r}^c_{il}} \sum_{l \in \mathscr{H} }  \frac{1}{{\sigma_l}^2} (\hat{r}^c_{il}-x)^2
    \end{equation}

      %However, $ \max_{l} \hat{r}^c_{il}-\min_{l} \hat{r}^c_{il}  \leq 2\delta$, therefore (\ref{mle1}) can be transformed as,

%         \begin{equation}
%    \label{mle2}
%   \mathcal{MLE}({{\hat{r}^c}_i}) = \arg \min_{ \min_{l} \hat{r}^c_{il}  \leq x \leq \max_{l} \hat{r}^c_{il}} \sum_{l=1}^L  \frac{1}{{\sigma_l}^2} {(x-\hat{r}^c_{il})}^2
 %   \end{equation}
%After brief calculation on the derivative of (\ref{mle1}), %$\frac{ d \sum_{l=1}^L  \frac{1}{{\sigma_l}^2} {(x-\hat{r}^c_{il})}^2 }{dx}$, equals
     %  \begin{equation}
   %    \begin{aligned}
 %  \sum_{l=1}^{L} \frac{1}{{\sigma_l}^2} 2 {(x-\hat{r}^c_{il})} = (\sum_{l=1}^L  \frac{2}{{\sigma_l}^2})  x -  \sum_{l=1}^L  \frac{2\hat{r}^c_{il}}{{\sigma_l}^2}
%    \end{aligned}
 %   \end{equation}
%    where ${ \min_{l} \hat{r}^c_{il}  \leq x \leq \max_{l} \hat{r}^c_{il}}$.
    It is easy to get the conclusion that the right hand of (\ref{mle1}) is minimized when $x=\frac{ \sum_{l \in \mathscr{H} }  \frac{\hat{r}^c_{il}}{{\sigma_l}^2}} {\sum_{l \in \mathscr{H}} \frac{1}{{\sigma_l}^2}}$ and thus
    \begin{equation}
    \label{MLEclosed}
      \mathcal{MLE}({{\hat{r}^c}_i}) = \frac{ \sum_{l \in \mathscr{H}}  \frac{\hat{r}^c_{il}}{{\sigma_l}^2}} {\sum_{l \in \mathscr{H}}  \frac{1}{{\sigma_l}^2}}
      \end{equation}
With the closed form proposed, the complexity of  determining $ \mathcal{MLE}({{\hat{r}^c}_i})$ is reduced to $O(1)$. %Though $\hat{r}^c_{il}$ here are obtained following ()-() for any $N$, while in essence we have converted determining the MLE of $\hat{r}^c_i$, $i=1,2,...,N,$ into $N$ separate steps via (\ref{MLEclosed}).
   % The method proposed here can be directly applied in \cite{2015mle}. Here we give more details to show why we develop such a definition of $\hat{r}^c_{il}$. Denote $\widetilde{r}^{c}_{i\epsilon(l)}$ as an ascending order of  $\widetilde{r}^{c}_{il}$ where $\epsilon(l), l=1,2,...,L,$ is a permutation of $\{1,2,...,L\}$. Due to $d_{\Gamma}(\widetilde{r}^{c}_{i\epsilon(L)}, \widetilde{r}^{c}_{i\epsilon(1)}) \leq 2\delta <\frac{\Gamma}{4}$, there are only two cases,
%   \begin{itemize}
 %  \item(1) $\widetilde{r}^{c}_{i\epsilon(L)}-\widetilde{r}^{c}_{i\epsilon(1)} < 2\delta$, shown in Figure \ref{rel}(a).
 %  \item(2) $\widetilde{r}^{c}_{i\epsilon(L)}-\widetilde{r}^{c}_{i\epsilon(1)} > 2\delta$,  shown in Figure \ref{rel}(b).
%   \end{itemize}

%    For Case (1), $\hat{r}^c_{il}$ is defined as (61). As for Case (2), when $\widetilde{r}^{c}_{il} \in [0,2\delta)$,  $\hat{r}^c_{il}$ is defined as (62); when $\widetilde{r}^{c}_{il} \in [\Gamma-2\delta, \Gamma)$,  $\hat{r}^c_{il}$ is defined as (63). In Figure \ref{rel}(c), we explain the definition of $\hat{r}^c_{il}$ with a geometry illustration, which is a shift of one side near $\Gamma$ to the left side of $0$. Therefore, the minimizing of (\ref{mle}) under $d_{\Gamma}$ is converted into optimization in pure Euclid distance in (85). \

%\begin{thebibliography}{00}


%%\bibitem{}

%\end{thebibliography}
\section{Simulation Results}
\noindent We have shown that how to generalize conventional CRT to solve the ambiguity resolution problems. The most ideal estimation we can expect is that the reconstruction error is linear to the residue error, since when $m_l > X_i$, the samples $\widetilde{r}_{il}$ should be of $X_i$ itself. In the following simulation, a robust estimation is defined as that $|\widetilde{X}_i - X_i| \leq 3 \frac{\Gamma}{4N}$ are satisfied for $i=1,2,...,N$. We assume that $\Delta_{il}$ are independent and identically distributed and follow a normal  distribution $N(0,\sigma^2)$ where $SNR = -20 \log_{10} \sigma$.

\begin{figure}
\centering
\includegraphics[width=2.6 in,height=2 in]{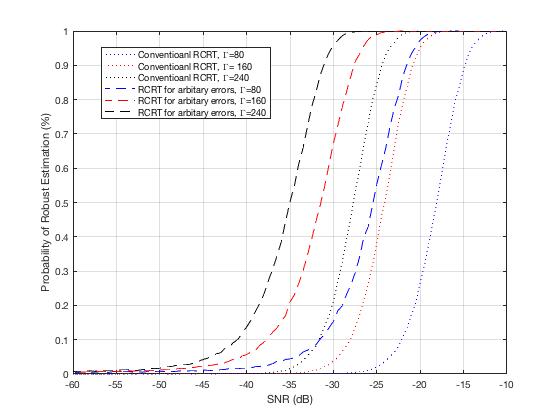}
\caption{Performance simulation comparison when $N=2$}
\label{sim1}
\end{figure}

\begin{figure}
\centering
\includegraphics[width=2.6 in,height=2 in]{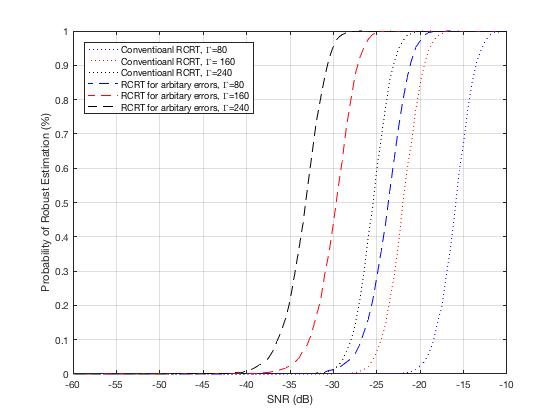}
\caption{Performance simulation comparison when $N=3$}
\label{sim2}
\end{figure}

In the simulation, we set $N=2, K=4, L=6$ and $N=3, K=6, L=10$ respectively where SNR is ranged from $-60$dB to $-10$dB. The results are shown in Fig \ref{sim1} and \ref{sim2}, which verify that the proposed scheme bring considerable improvement in strengthening the robustness.

\section{Conclusion}
\noindent In this paper, the first robust Chinese Remainder Theorem tolerating arbitrary errors for multiple numbers has been proposed. Various optimizations have been developed to both reduce the computational complexity and  improve the robustness performance to further widen applications of RCRTMN.

%\endinput
%%
%% End of file `elsarticle-template-harv.tex'.

%\ifCLASSOPTIONcaptionsoff
%  \newpage
%\fi
%\bibliographystyle{plain}
%\bibliography{reference.bib}

%It is not necessary to upload the biography when you submit your manuscript.

\end{document}